\documentclass{ws-ijmpd}

\usepackage[super,compress]{cite}
\usepackage{epsfig}
\usepackage{graphicx}
\usepackage{amsfonts,amssymb,amsmath}

\newcommand{\Tr}{\mbox{Tr}}

\begin{document}



\title{On the flat spacetime Galileons and the Born-Infeld type structures}

\author{CUAUHTEMOC CAMPUZANO}

\address{Instituto de F\'\i sica, Pontificia Universidad Cat\'olica
de Valpara\'\i so
\\
Casilla 4950, Valpara\'\i so, Chile
\\
and
\\
Facultad de F\'\i sica, Universidad Veracruzana
\\
91000 Xalapa, Veracruz, M\'exico
}

\author{RUB\'EN CORDERO}

\address{Departamento de F\'\i sica, Escuela Superior de F\'\i sica 
y Matem\'aticas del I.P.N. \\
Unidad Adolfo L\'opez Mateos, Edificio 9, 07738 M\'exico, 
Distrito Federal, M\'exico
}

\author{MIGUEL CRUZ}

\address{Instituto de F\'\i sica, Pontificia Universidad Cat\'olica
de Valpara\'\i so
\\
Casilla 4950, Valpara\'\i so, Chile
}

\author{EFRA\'IN ROJAS}

\address{Facultad de F\'\i sica, Universidad Veracruzana
\\
91000 Xalapa, Veracruz, M\'exico
}

\maketitle



\begin{abstract}
We show how the flat spacetime Galileon field theories (FSGFT) in arbitrary dimensions
can be obtained through a Born-Infeld type structure. This construction involves 
a brane metric and non-linear combinations of derivatives of a scalar field. Our 
setup gives rise to some Galileon tensors and vectors useful for the variational 
analysis which are related to the momentum density of the probe Lovelock branes 
floating in a $N$-dimensional flat bulk. We find further that the Noether currents 
associated to these Galileon theories may be written in terms of such tensors.

\end{abstract}

\ccode{PACS Nos.:04.20.Fy, 04.50.Kd, 10.10Kk, 11.25.-w, 98.80.Cq}

\section{Introduction}
\label{intro}
%

Recently there has been a lot of interest in Galileon field theories because they
may have implications for particle physics and cosmology. They have the 
ability to produce an accelerated expansion scenario in the absence of any type 
of matter interaction~\cite{galileon,rham1,trodden5,trodden2,trodden} as well as to 
exhibit the Vainshtein screening mechanism~\cite{vainshtein,deffayet-1,deffayet-0}
at short distances. On theoretical grounds, these scalar field models with derivative 
self-interactions are second-order derivative theories which, surprisingly, are 
completely healthy non-higher derivative theories because their equations of motion 
remain second-order. They have been covariantized and extended to $p$-forms~\cite{deser1}.
Besides, these theories are anchored to the former Horndeski's scalar-tensor theories in 
curved 4-dimensional spacetimes~\cite{horn,gali-horn} and were discovered 
independently for flat spacetimes by Fairlie~\cite{fairlie}. 

Following a geometric viewpoint the Galileons can be derived from the perspective of a
brane probing a background spacetime as was developed in~\cite{rham1,trodden2,trodden,trodden3,trodden4}. 
In such brane prescription Galileons follow from the Lovelock invariants defined on 
the worldvolume swept out by a spacelike brane \cite{rham1,lovelock,bil,van}, 
evaluated in an unitary gauge. For a $N$-dimensional background spacetime an interesting 
subclass of these theories is provided when such bulk is flat Minkowski, ${\cal M}_N$. There are 
two possible foliations for the ${\cal M}_N$ viewed as a maximally symmetric space by 
$(N-1)$-dimensional maximally symmetric timelike slices; that is, ${\cal M}_N$ can be 
foliated by flat ${\cal M}_{N-1}$ slices or by $dS_{N-1}$ slices, namely. Such flat 
spacetime Galileon field theories (FSGFT) are known as DBI Galileons and type II $dS$ 
DBI Galileons, respectively~\cite{rham1,trodden2,trodden,trodden6,trodden7}. Another 
constructions are possible but they are in dependence on the curved nature of the background 
spacetime~\cite{trodden2,trodden,trodden8}. 

In this paper it is shown how the flat spacetime Galileon field theories can 
be obtained from a Born-Infeld (BI) type action. This action is in connection 
with a Born-Infeld-Lovelock (BIL) framework developed for describing Lovelock
brane models~\cite{bil} and is written close in the spirit to the one developed 
in~\cite{deser}. Our construction contains contribution from a brane metric and 
non-linear combinations of the derivatives of the scalar field defined on the brane. 
Despite the bulk is Minkowski this is a case of sufficient complexity that still 
deserves further explorations. By expanding this determinantal BI Lagrangian in terms 
of traces one finds a finite series containing all the Galileon terms for a given 
dimension. We may, additionally, make a variational analysis in order to obtain the 
equations of motion (eom) and to study the associated Noether currents.

The paper is organized as follows. The aim of Section 2 is to acquaint the reader
with basic covariant facts of the FSGFT. In Section 3 we introduce a BI type action 
which, when expanded, contains all the FSGFT for any arbitrary dimension. This number
is in dependence of the dimension of the worldvolume. In addition, we provide another 
BI-like structures pursuing the same claim. In Section 4 we introduce some Galileon 
tensors and vectors useful to express the Noether currents associated to these models 
and discuss some of their properties. Section 5 is devoted to the forthright derivation 
of the equations of motion and the Noether currents. We conclude in Section 6 with 
some comments. Appendices gather information about the brane geometry in the unitary 
gauge and some mathematical relations useful for expanding the BI-like structures. The 
notation used in the paper is the usual one. When working with the scalar field $\phi$ 
in curved spacetime with metric $q_{ab}$ and covariant derivative $\nabla_a$, we use 
the notation $\pi_a = \nabla_a \phi$ and $\Pi_{ab} := \nabla_a \nabla_b \phi$. For trace 
of powers of the matrix $\Pi_{ab}$ we have $[\Pi^n] := \Tr (\Pi^n)$, e. g., $[\Pi] = 
\nabla^a \nabla_a \phi$, $[\Pi^2] = \nabla_a \nabla_b \phi \,\nabla^a \nabla^b \phi$ 
where the brane indices $a,b$ are raised with the inverse metric $q^{ab}$. The contractions 
of powers of $\pi$ with $\Pi$ are usually denoted by $[\pi^n] = \pi\cdot \Pi^{n-2} \cdot 
\pi$, e. g., $[\pi^2] = \pi_a \pi^a$, $[\pi^3] = \pi_a \Pi^{ab} \pi_b$.

%
\section{Flat spacetime Galileons in arbitrary dimension}
\label{sec:galileons}
%

Consider the action governing the dynamical evolution of a single scalar field $\phi(x^a)$ 
living in an orientable codimension one brane, $\Sigma$, with local coordinates $x^a$ and 
metric $q_{ab}$ $(a,b = 0,1,\ldots,p)$ embedded in a $N= (p+2)$-dimensional Minkowski 
background spacetime with metric $\eta_{\mu \nu}$ $(\mu , \nu = 0,1,2, \ldots, p+1)$  
\begin{equation}
S[\phi] =  \int_\Sigma d^{p+1}x \, \sqrt{-q} \sum_{n=0} ^{p} \alpha_n\, L_n
(\phi,\nabla_a \phi, \nabla_a\nabla_b \phi),
\label{eq:LbGalileon}
\end{equation}
where
\begin{equation}
L_n = \gamma^{-1} f^{p} \,\delta^{a_1 a_2 a_3 \cdots a_n} _{b_1 b_2 b_3 \cdots b_n}
h^{b_1}{}_{a_1} h^{b_2}{}_{a_2} h^{b_3}{}_{a_3} \cdots h^{b_n}{}_{a_n},
\label{eq:lovelock-brane-again}
\end{equation} 
and $\delta^{a_1 a_2 a_3 \cdots a_n} _{b_1 b_2 b_3 \cdots b_n}$ being the 
generalized Kronecker delta (gKd), $q=\mbox{det} \,(q_{ab})$ and $\gamma = 
(1+ f^{-2} \nabla^a \phi \nabla_a \phi)^{-1/2}$. Here and in what follows, 
$f=f(\phi)$ such that $f=1$ for the DBI Galileons whereas $f=\phi$ for the 
type II $dS$ DBI Galileons. In Planck units the coupling constants $\alpha_n$ 
have dimensions $[\alpha_n] = L^{n-p-1}$. Moreover, we have introduced the matrix
\begin{equation}
h_{ab} := \gamma f^{-2} \left[ -\Pi_{ab} + f^{-2} \gamma^2 \left( \pi_{a}\Pi^c{}_b
\pi_c + f f' \pi_{a} \pi_b \right) + ff' q_{ab} \right],
\label{eq:Hmatrix}
\end{equation}
with $f' = \partial f/\partial \phi$. The brane indices $a,b$ are raised and lowered
with $q^{ab}$ and $q_{ab}$, respectively. Only the first $n$ of these Galileon
Lagrangians are non-trivial in $n \leq p+1$ dimensions. In addition, $\nabla_a$ denotes 
the covariant derivative compatible with $q_{ab}$. It is worth commenting on the origin of the
matrix (\ref{eq:Hmatrix}). This arise from the contraction between the inverse of the induced 
metric $g^{ab}$ and the extrinsic curvature $K_{ab}$ of the $\Sigma$ brane when they are 
expressed in terms of the unitary gauge~(see~(\ref{eq:induced-1}) and (\ref{eq:KabG})).
Thus, $h^a{}_b = g^{ac}K_{cb}$. In connection with this fact the brane induced metric
on $\Sigma$ is
\begin{equation}
g_{ab} = f^2 q_{ab} + \pi_a \pi_b,
\label{eq:gab}
\end{equation}
such that $g^{ac}g_{cb} = \delta^a{}_b$.

With regards the structure (\ref{eq:lovelock-brane-again}), we set $L_0 = \gamma^{-1} 
f^{p+1}$. Another elegant framework for obtaining other type of Galileon field theories 
using the anti-symmetric Levi-Civita symbol is provided in~\cite{deser1}. Geometrically 
speaking, $\Sigma$ represents the worldvolume swept out by a dynamical Lovelock $p$-brane 
which can be described locally by $y^\mu = X^\mu(x^a)$ evaluated in an unitary gauge where 
$y^\mu$ are local coordinates for the bulk and $X^\mu$ are the embedding functions. 
By expanding out Eq.~(\ref{eq:lovelock-brane-again}) in terms of the traces associated 
to the matrix (\ref{eq:Hmatrix}) (see~\ref{appC} for details on the traces of the $h_{ab}$
matrix), at the first few orders we have
\begin{eqnarray}
L_0 &=&\gamma^{-1} f^{p+1},
\label{eq:L0G}
\\
L_1 &=&\gamma^{-1} f^{p+1}\Tr(h), 
\nonumber
\\
&=& f^{p - 1} \left\lbrace - [\Pi] + \frac{\gamma^2}{f^2} [\pi^3]
+ ff' (p+2 - \gamma^2) \right\rbrace ,
\label{eq:L2G}
\\
L_2 &=& \gamma^{-1} f^{p+1}\, [\Tr^2 (h) - \Tr(h^2) ],
\nonumber
\\
&=& \gamma \,f^{p - 3} \left\lbrace [\Pi]^2 - [\Pi^2]
+ 2 \frac{\gamma^2}{f^2} \left( - [\Pi] [\pi^3] + [\pi^4] \right) 
\right.
\nonumber
\\
&+& \left. 2 f f' \left[ - (p+1) [\Pi] + \frac{\gamma^2}{f^2}\left( f^2 [\Pi] 
+ (p +1) [\pi^3] \right)  \right] \right.
\nonumber
\\
&+& \left. f^2 f^{'2} p (p+3 - 2 \gamma^2)  \right\rbrace ,
\\
L_3 &=& \gamma^{-1} \,f^{p+1}  [ \Tr^3 (h) - 3 \Tr(h) 
\Tr(h^2) + 2 \Tr(h^3) ],
\nonumber
\\
&=& \gamma^2 f^{p - 5} \left\lbrace - \left( [\Pi]^3 - 3 [\Pi]
[\Pi^2] + 2 [\Pi^3] \right) 
\right. 
\nonumber
\\
&-& \left. 3 \frac{\gamma^2}{f^2} \left[ - [\pi^3] \left( [\Pi]^2 - [\Pi^2]\right) 
+ 2 [\Pi] [\pi^4] - 2 [\pi^5] \right]  
\right. 
\nonumber
\\
&+& \left. 3 f f' \left[ p \left( [\Pi]^2 - [\Pi^2]\right) - \frac{\gamma^2}{f^2}
\left( f^2  \left( [\Pi]^2 - [\Pi^2]\right) + 2p \left( [\Pi] [\pi^3] 
- [\pi^4] \right)  \right) \right] 
\right.
\nonumber
\\
&-& \left. 3 f^2 f^{' \,2}(p-1) \left[ (p+2) [\Pi] - \frac{\gamma^2}{f^2} 
\left( 2f^2 [\Pi] + (p+2) [\pi^3] \right) \right]
\right.
\nonumber
\\
&+& \left. f^3 f^{'\,3}\,p\,(p-1)\,(p+4-3\gamma^2) \right\rbrace,
\label{eq:L4G}
\end{eqnarray}
where we have considered that $[\pi^2] = \pi_a \pi^a = f^2 \gamma^{-2} (1 - \gamma^2)$. 
Regarding the so-called {\it tadpole terms}, ${L}_T$, they are not constructed 
in the present way but as the $N$-dimensional proper volume bounded by the worldvolume 
(see the discussion in \cite{rham1,trodden2}). The equations of motion derived from 
any of these Lagrangians will contain no more than two derivatives on the scalar field, 
ensuring that no extra degrees of freedom propagate around the background spacetime as 
we will show below.

%
\section{Flat spacetime Galileons from Born-Infeld type structures}
%

The setup outlined above allows us to find another general expression 
involving the flat spacetime Galileon field theories.
Consider now the local Born-Infeld type action
\begin{equation}
\label{eq:action21}
S[\phi] =  \int_\Sigma d^{p+1}x \,\gamma^{-1}f^{p+1}
\sqrt{- \det (q_{ab} + X_{ab})}, 
\end{equation}
where 
\begin{equation}
X_{ab} :=  2 \alpha \,h_{ab} + \alpha^2\, h_{ac} h^c{}_b,
\label{eq:Xab}
\end{equation}
and $h_{ab}$ being defined in (\ref{eq:Hmatrix}) and $\alpha$ is a concomitant constant 
characterizing the relative weight of the nonlinear terms in this model. 

The BI type volume element form in (\ref{eq:action21}) may be written in terms of 
the flat spacetime Galileon Lagrangians for a given dimension $p$. Indeed, it follows 
from the fact that the combined matrix $G_{ab}:= q_{ab} + X_{ab}$ can be written as 
${G}_{ab}= q_{cd} \left( q^c{}_a{} +  \alpha h^c{}_a)(q^d{}_b + \alpha h^d{}_b\right)$. 
This entails that the action~(\ref{eq:action21}) can be re-expressed as
\begin{equation}
S[\phi] =  \int_\Sigma d^{p+1}x\, \sqrt{-q} \,\gamma^{-1} f^{p+1}
\,\left[ \textrm{det} \left( q^a{}_b + \alpha\,h^a{}_b \right)\right].  
\label{eq:action23}
\end{equation}
Now we turn to expand the characteristic polynomial of $h^a{}_b$ by using the identity 
(\ref{eq:det-2}). To do this we need once again the traces of $h_{ab}$ (see \ref{appC}). 
It therefore follows, from Eqs. (\ref{eq:X1}-\ref{eq:X4}) specialized to $f_{ab} = \alpha\, 
h_{ab}$, that after a lengthy but straightforward computation that
\begin{eqnarray}
f_{(1)} &=& \alpha f^{p - 1} \left[ - [\Pi] + \frac{\gamma^2}{f^2} [\pi^3]
+ ff' (p+2 - \gamma^2) \right],
\label{eq:f1}
\\
f_{(2)} &=& \frac{\alpha^2}{2}\gamma \,f^{p - 3} \left\lbrace [\Pi]^2 - [\Pi^2]
+ 2 \frac{\gamma^2}{f^2} \left( - [\Pi] [\pi^3] + [\pi^4] \right) 
\right.
\nonumber
\\
&+& \left. 2 f f' \left[ - (p+1) [\Pi] + \frac{\gamma^2}{f^2}\left( f^2 [\Pi] 
+ (p +1) [\pi^3] \right)  \right] \right.
\nonumber
\\
&+& \left. f^2 f^{'\,2} \,p\, (p+3 - 2 \gamma^2)  \right\rbrace ,
\\
f_{(3)} &=& \frac{\alpha^3}{6}\gamma^2 \,f^{p - 5} \left\lbrace - \left( [\Pi]^3 - 3 [\Pi]
[\Pi^2] + 2 [\Pi^3] \right) 
\right. 
\nonumber
\\
&-& \left. 3 \frac{\gamma^2}{f^2} \left[ - [\pi^3] \left( [\Pi]^2 - [\Pi^2]\right) 
+ 2 [\Pi] [\pi^4] - 2 [\pi^5] \right]  
\right. 
\nonumber
\\
&+& \left. 3 f f' \left[ p \left( [\Pi]^2 - [\Pi^2]\right) - \frac{\gamma^2}{f^2}
\left( f^2  \left( [\Pi]^2 - [\Pi^2]\right) + 2p \left( [\Pi] [\pi^3] 
- [\pi^4] \right)  \right) \right] 
\right.
\nonumber
\\
&-& \left. 3 f^2 f^{' \,2}(p-1) \left[ (p+2) [\Pi] - \frac{\gamma^2}{f^2} 
\left( 2f^2 [\Pi] + (p+2) [\pi^3] \right) \right]
\right.
\nonumber
\\
&+& \left. f^3 f^{'\,3}\,p\,(p-1)\,(p+4-3\gamma^2) \right\rbrace,
\\
& \vdots &
\nonumber
\\
f_{(s)} &=& \frac{\alpha^s}{s!} L_s,
\label{eq:fs}
\end{eqnarray}
where $L_s$ is given by Eq.~(\ref{eq:lovelock-brane-again}). If we set $f_{(0)} =1$
and then substitute all these results in (\ref{eq:det-2}) and therefore into the 
action~(\ref{eq:action23}) we obtain that the BI type action~(\ref{eq:action21}) 
can be expressed as 
\begin{equation}
S[\phi] =  \int_\Sigma d^{p+1} x\, \sqrt{-q} \sum_{n=0} ^{p} \left( \frac{\alpha^n}{n!} 
\right) L_{n},
\label{eq:action4}
\end{equation}
where the Galileon Lagrangian functions (\ref{eq:L0G}-\ref{eq:L4G}) have been invoked
and so on. It should be stressed that this action is similar to the action~(\ref{eq:LbGalileon})
whenever $\alpha_n = \alpha^n / n!$. This structure is a particular 
case of the action (\ref{eq:LbGalileon}). With regards this point, a non-trivial question
is how to find the correct couplings $\alpha_n$ that yield a viable physical theory but, 
for the moment, such point is beyond the scope of this work. An interesting proposal
on this subject relating the unitary analysis of BI gravity actions is developed 
in~\cite{tekin2}. Further, as in the gravitational case, one could relate this type of 
BI-like actions to a Chern-Simons limit of the Galileon theory by a particular choice of 
the couplings $\alpha_n$ in the series (\ref{eq:LbGalileon})~\cite{zanelli}. Hence, for 
our purposes, when the flat spacetime Galileon field theories are configured as 
in~(\ref{eq:LbGalileon}), these admit a Born-Infeld type structure.

\subsection{Other BI-like structures}

The combination $X_{ab}$ is not constrained to have the form~(\ref{eq:Xab}) but it  
seems reasonable to think of some other possible choices for $X_{ab}$ in order to
reproduce the flat spacetime Galileon Lagrangian terms. Some of them are more complex
while others are limited but even so all of them contain interesting information.
Consider for example
\begin{equation}
X_{ab} = \left( \alpha + \beta\,h \right) h_{ab},
\label{eq:Xab2}
\end{equation}
where $h= \Tr (h_{ab})$ and $\beta$ being another constant. The action~(\ref{eq:action21}) 
can be rewriten as
\begin{equation}
S[\phi] =  \int_\Sigma d^{p}x\, \sqrt{-q} \gamma^{-1}f^{p+1}
\sqrt{ \det ( \delta^a{}_b + X^a{}_b ) }.
\label{eq:action31}
\end{equation}
By using the expansion~(\ref{eq:expansion}) we note that if $\beta = \alpha^2 / 4$ then
\begin{eqnarray}
\sqrt{ \det (\delta^a{}_b + X^a{}_b)} &=&
1 + \frac{\alpha}{2} h + \frac{\alpha^2}{4} \left[ h^2 - \Tr (h^2) 
\right] + \frac{\alpha^3}{12} \left[ h^3 - 3 h\Tr (h^2) + 2
 \Tr (h^3) \right]
\nonumber
\\
&+& \frac{\alpha^4}{192} \left[ 5h^4 - 27 h^2 \Tr (h^2)  
+ 40 h \Tr (h^3) + 6 \Tr (h^2)^2 - 24   \Tr (h^4) 
\right] 
\nonumber
\\
&+& {\cal O} (h^5).
\end{eqnarray}
When we substitute this expression into the action~(\ref{eq:action31}) by
considering the matrix~(\ref{eq:Hmatrix}), we obtain
\begin{equation}
S[\phi] =  \int_\Sigma d^{p+1}x\, \sqrt{-q} \left[ 
L_0 + \frac{\alpha}{2}L_1 + \frac{\alpha^2}{4} L_2 + \frac{\alpha^3}{12} L_3
+ {\cal O}(h^4)\right].
\end{equation}
In other words, in the small value of $X_{ab}$ given by~(\ref{eq:Xab2}) the action
(\ref{eq:action31}) casts out only the first four terms of the Galileon Lagrangians 
but not beyond this. Again, we have invoked the expressions~(\ref{eq:L0G}-\ref{eq:L4G}) 
and so on. This action might be useful if one is interested only in $(3+1)$ dimensions. 
In addition, the choice $X_{ab} = \alpha\,h_{ab}$ results very limited because 
the corresponding expansion for small values of $X_{ab}$ only reproduce the first 
two Galileon Lagrangian terms.

%
\section{Galileon tensors}
\label{sec:tensors}
%

For future reference we introduce the following tensors for each $n$ value
\begin{equation}
J^{ab} _{(n)} := \delta^{aa_1 a_2 \cdots a_n} _{b_0 b_1 b_2 \cdots b_n}
g^{bb_0} h^{b_1}{}_{a_1} h^{b_2} {}_{a_2} \cdots h^{b_n} {}_{a_n},
\label{eq:Jab}
\end{equation}
which are symmetric. This fact can be proved with the aid of the gKd properties and 
the definition $h^a{}_b = g^{ac}K_{cb}$. By expanding out this expression in terms of 
minors we have a recursion relation
\begin{eqnarray}
J^{ab} _{(n)} &=& \left[ \delta^a _{b_0} \delta^{a_1a_2a_3\cdots a_n} _{b_1b_2b_3 \cdots b_n}
-  \delta^a _{b_1} \delta^{a_1a_2a_3\cdots a_n} _{b_0b_2b_3 \cdots b_n} 
+ \cdots + (-1)^n \delta^a _{b_n} \delta^{a_1a_2a_3\cdots a_n} _{b_0b_1b_2 \cdots b_{n-1}} 
\right]\times
\nonumber 
\\ 
&&\hspace{7cm}  g^{bb_0} h^{b_1}{}_{a_1} h^{b_2}{}_{a_2}\cdots h^{b_n} {}_{a_n}, 
\nonumber
\\
&=& \gamma f^{-(p+1)} L_n g^{ab} - n h^a{}_c J^{cb} _{(n-1)},
\label{eq:recursion-1}
\end{eqnarray}
where we have used the expression defining the Lagrangians~(\ref{eq:lovelock-brane-again})
and the definition~(\ref{eq:Jab}). In addition, when we contract the expression~(\ref{eq:Jab}) 
with the extrinsic curvature~(\ref{eq:KabG}) we obtain the important identity
\begin{eqnarray}
J^{ab} _{(n)} K_{ab} &=& \gamma f^{-(p+1)} L_{n+1},
\label{eq:id-1}
\end{eqnarray}
where we have used the relation~(\ref{eq:lovelock-brane-again}) again. Similarly, when we
contract the tensors (\ref{eq:Jab}) with the brane induced metric~(\ref{eq:gab})
we have the relation
\begin{equation}
J^{ab} _{(n)} g_{ab} = \gamma f^{-(p+1)} (p+1-n) L_n.
\label{eq:id-2}
\end{equation}
We will refer hereafter these tensors as {\it Galileon tensors}.

We introduce now the worldvolume vectors
\begin{equation}
J^a _{(n)} := \gamma^{-1} J^{ab} _{(n)} \pi_b.
\label{eq:Ja}
\end{equation}
Taking into account Eq.~(\ref{eq:recursion-1}) and the definition (\ref{eq:Ja}) we 
obtain a recursion relation for these vectors, $J^a _{(n)} = \gamma^2 f^{-(p+3)} L_n \pi^a 
- nh^a{}_b J^b _{(n-1)}$, where we have used expression~(\ref{eq:induced-1}) as well as 
expressing $[\pi^2]$ in favor of $\gamma^{-1} = (1 + f^{-2} [\pi]^2)^{1/2}$. A convenient 
form for this relation is obtained when we insert the definition of the $h_{ab}$ 
matrix, Eq. (\ref{eq:Hmatrix}), 
\begin{eqnarray}
J^a _{(n)} &=& \gamma^2 f^{-(p+3)} L_n  \,\pi^a + n \gamma f^{-2}  \Pi^a{}_b J^b _{(n-1)}
- n \gamma^3 f^{-4} \pi^a  \pi^b \Pi_{bc} J^c _{(n-1)} 
\nonumber
\\
&-& n \gamma^3 f^{-3} f' \pi^a \pi_b J^b _{(n-1)} - n \gamma f^{-1} f' J^a _{(n-1)}.
\label{eq:recursive-3}
\end{eqnarray}
These vectors are to be referred to as the {\it Galileon vectors}. For illustration,
in \ref{app:B} we give some values of the tensors (\ref{eq:Jab}) and the 
vectors~(\ref{eq:Ja}). 
Moreover, from Eqs. (\ref{eq:KabG}) and (\ref{eq:id-1}) we have
\begin{equation}
\gamma^{-1} f^{p+1} \Pi_{ab} J^{ab} _{(n)} = (p+1-n)f^{-1}f'L_n - \gamma^{-1}L_{n+1}
+ f^p f'\pi_a J^a _{(n)}.
\label{eq:id-4}
\end{equation}
Alike from Eqs. (\ref{eq:gab}) and (\ref{eq:id-2}) we have that the trace of the 
Galileon tensors is
\begin{equation}
\Tr (J^{ab} _{(n)}) = J^{ab} _{(n)} q_{ab} = (p+1-n) \gamma f^{-(p+3)} L_n - 
\gamma f^{-2} \pi_a J^a _{(n)}.
\label{eq:id-13}
\end{equation}

\noindent
The conservation of the Galileon tensors on regards the geometry provided by the 
induced metric~(\ref{eq:gab}) is ${\cal D}_a J^{ab} _{(n)} = 0$ where ${\cal D}_a$ 
is the covariant derivative compatible with $g_{ab}$ (see \cite{bil} for details). 
In terms of the geometry provided by $q_{ab}$ this relation becomes 
$\nabla_a (\gamma^{-1} J^{ab}_{(n)}) - f^{-2} \pi^b K_{ac} J^{ac} _{(n)} + \gamma^{-1}
f^{-1} f' (p+3) J^{ab}_{(n)} \pi_a = 0$ where once again we have made use of the formulae in
\ref{app:A}. Hence, by using the identity (\ref{eq:id-1}) and the definition of the 
Galileon vectors (\ref{eq:Ja}) we obtain 
\begin{equation}
 \nabla_a \left( \gamma^{-1} f^{p+1} J^{ab} _{(n)} \right) = \gamma f^{-2}L_{n+1} \pi^b
-  2 f^p f' J^{b} _{(n)}.
\label{eq:conservation-2}
\end{equation}
In this spirit, it remains to obtain a divergence expression for the Galileon vectors. 
The contraction of~(\ref{eq:conservation-2}) with $\pi_b$ and the use of Eq. (\ref{eq:Ja}) 
yields
\begin{equation}
\nabla_a (f^{p+1} J^a _{(n)}) = \gamma^{-1} (1-\gamma^2) L_{n+1} 
+ \gamma^{-1} f^{p+1} \Pi_{ab} J^{ab} _{(n)} - 2 f^p f' \pi_a J^a _{(n)}. 
\label{eq:divergence-3}
\end{equation}
Finally, by using Eqs. (\ref{eq:id-4}), (\ref{eq:divergence-3}) and (\ref{eq:id-13}) we have
another useful identity given by
\begin{equation}
\gamma L_{n+1} = - \nabla_a (f^{p+1} J^a _{(n)}) + \gamma^{-1} f^{p+2} f' \Tr (J^{ab} _{(n)}).
\label{eq:id-14}
\end{equation}

\subsection{Auxiliary Galileon tensors}

In analogy with the relation (\ref{eq:Jab}) and for convenience in the obtaining of the 
eom and the Noether currents it will be useful to introduce some auxiliary Galileon tensors
\begin{equation}
{\cal J}^a _{(n)b} := \delta^{a a_1 a_2 a_3 \cdots a_n} _{b b_1 b_2 b_3 \cdots b_n}
h^{b_1}{}_{a_1}h^{b_2}{}_{a_2}h^{b_3}{}_{a_3} \cdots h^{b_n}{}_{a_n}.
\label{eq:Jab-aux}
\end{equation}
In general these are not symmetric. Obviously ${\cal J}^a _{(n)b} = J^{ac} _{(n)} g_{cb}$.
In terms of the Galileon tensors and vectors these explicitly reads
\begin{equation}
{\cal J}^{ab} _{(n)} = f^2 J^{ab} _{(n)} + \gamma J^a _{(n)} \pi^b.
\label{eq:id-11}
\end{equation}
Finally, by construction these tensors satisfy
\begin{equation}
{\cal J}^{ab} _{(n)} h_{ba} = \gamma f^{-(p+1)} L_{n+1}.
\label{eq:id-12}
\end{equation}

%
\section{Equations of motion and Noether currents in arbitrary dimensions}
%

Let us consider an infinitesimal deformation $\phi \to \phi + \delta \phi$ of the
scalar field describing the normal deformation for the worldvolume $\Sigma$. Under
this deformation the basic quantities that characterize $\Sigma$ change according
to 
\begin{eqnarray}
\delta \pi_a &=& \nabla_a \delta \phi,
\label{eq:var2}
\\
\delta \Pi_{ab} &=& \nabla_a \nabla_b \delta \phi,
\label{eq:var3}
\\
\delta \gamma &=& f^{-1} f' \gamma (1- \gamma^2)\delta \phi
- \gamma^3 f^{-2} \pi^a \nabla_a \delta \phi,
\label{eq:var4}
\\
\delta h_{ab} &=& - \gamma^2 f^{-1} f' \left[ h_{ab} + \nabla_b (\gamma f^{-2} \pi_a) -
2\gamma^{-1} f^{-3}f' \pi_a \pi_b \right] \delta \phi
\nonumber
\\
&-& \gamma^2 f^{-2} \left[ h_{ab} + \nabla_b (\gamma f^{-2} \pi_a) \right]\pi^c \nabla_c \delta \phi
- 2 \gamma f^{-3} f' \pi_{(a} \nabla_{b)} \delta \phi
\nonumber
\\
&-& \nabla_b \delta (\gamma f^{-2} \pi_a),
\label{eq:var5}
\end{eqnarray}
where in addition we have considered that $\delta f = f' \delta \phi$ and $\delta f' =0$. Using these 
expressions the variation of the Galileon action (\ref{eq:action4}) reads
\begin{eqnarray} 
\delta S [\phi] &=& \int_\Sigma dV \sum_{n=0} ^{p} \beta_n \left[ 
\gamma L_n  \delta \gamma^{-1} + (p+1) f^{-1}f'\,L _n\,\delta \phi 
+  n \gamma^{-1} f^{p+1} {\cal J}^{ab} _{(n-1)} \delta  h_{ba} \right],
\nonumber
\\
&=& \int_\Sigma dV \sum_{n=0} ^{p} \beta_n \left[ 
f^{-1}f'(p+\gamma^2)\,L _n\,\delta \phi + \gamma^2 f^{-2}L _n\,\pi^a
\nabla_a \delta \phi \right. 
\nonumber
\\
& & \hspace{2cm} \left. + n \gamma^{-1} f^{p+1} {\cal J}^{ab} _{(n-1)} \delta h_{ba} \right],
\nonumber
\end{eqnarray}
where we have written $dV:= d^{p+1}x \sqrt{-q}$ and $\beta_n :=  \alpha_n / n!$ 
for the sake of brevity. Taking into account (\ref{eq:var5}) and (\ref{eq:id-12}) the 
variation of the action reads
\begin{eqnarray}
\delta S &=& \int_\Sigma dV \sum_{n=0} ^{p} \beta_n \left\lbrace 
f^{-1}f'(p+\gamma^2) L_n \delta \phi + \nabla_a \left( \gamma^2 f^{-2} L_n \pi^a \delta \phi \right)
\right.
\nonumber
\\
&-& \left. 
nf^{p-2}f' {\cal J}^{ba} _{(n-1)} \pi_b \nabla_a \delta \phi
-  n \gamma^2 f^{-1}f' L_n \delta \phi - n\gamma^2 f^{-2} L_n \pi^a \nabla_a \delta \phi
\right. 
\nonumber
\\
&+& \left. 
2n f^{p-1} f^{'2} \gamma^{-1} J^a _{(n)} \pi_a \delta \phi 
-  n\gamma^{-1} f^p f' J^a _{(n-1)} \nabla_a \delta \phi 
\right. 
\nonumber
\\
&-& \left.  
n \gamma f^p f' {\cal J}^{ab} _{(n-1)} \nabla_a (\gamma f^{-2} \pi_b ) \delta \phi
- n \gamma f^{p-1} {\cal J}^{bc} _{(n-1)} \nabla_b (\gamma f^{-2} \pi_c) \pi^a \nabla_a \delta \phi
\right.
\nonumber
\\
&-& \left. 
\nabla_a \left( \gamma^2 f^{-2} L_n  \pi^a \right) \delta \phi 
- \nabla_a \left[n\gamma^{-1}f^{p+1} {\cal J}^{ab} _{(n-1)} \delta (\gamma f^{-2} \pi_b ) \right]
\right. 
\nonumber
\\
&+& \left. 
\nabla_a \left[n\gamma^{-1}f^{p+1} {\cal J}^{ab} _{(n-1)}\right] \delta (\gamma f^{-2} \pi_b )
\right\rbrace.
\nonumber
\end{eqnarray}

\noindent
After integrations by parts and considering the identities and definitions provided 
by Section \ref{sec:tensors} a rather long but straightforward computation leads to
\begin{equation}
\delta S [\phi] = \int_\Sigma dV \sum_{n=0} ^{p}  
\beta_n \left[ {\cal E}_n \delta \phi + \nabla_a Q^a _{(n)}(\delta \phi, 
\nabla_b \delta \phi ) \right] ,
\label{eq:var-3}
\end{equation}
where
\begin{eqnarray}
{\cal E}_n &=& f^{-1} f' (p+\gamma^2)L_n  
- \gamma (1+\gamma^2) f^{-3} f' \pi_b \nabla_a (n\gamma^{-1} f^{p+1} {\cal J}^{ab} _{(n-1)})
\nonumber
\\
&+& 2n \gamma^{-1} f^{p-1} f^{'2} \pi_a J^a _{(n-1)} - n\gamma f^p f' {\cal J}^{ab} _{(n-1)} \nabla_a (\gamma f^{-2} \pi_b)  
- n \gamma^2 f^{-1} f' L_n
\nonumber
\\
&+& \nabla_a \left\lbrace - \gamma^2 f^{-2} \pi^a L_n + n\gamma^2 f^{-2} \pi^a L_n 
+ n\gamma f^{p-1} {\cal J}^{bc} _{(n-1)} \nabla_b (\gamma f^{-2} \pi_c ) \pi^a 
\right. 
\nonumber
\\
&+& \left. n \gamma^{-1} f^p f' J^a _{(n-1)} +  \gamma^3 f^{-4} \pi^a \pi_b 
\nabla_c (n \gamma^{-1} f^{p+1} {\cal J}^{cb} _{(n-1)}) 
\right. 
\nonumber
\\
&-& \left. \gamma f^{-2} \nabla_b (n\gamma^{-1} f^{p+1} {\cal J}^{ba} _{(n-1)}) 
+ nf^{p-2} f' {\cal J}^{ba} _{(n-1)} \pi_b 
\right\rbrace,
\label{eq:eom-1}
\end{eqnarray}
and 
\begin{eqnarray}
Q_{(n)} ^a &=& \left[  \gamma^2 f^{-2} L_n \pi^a - n \gamma^2 f^{-2}  L_n \pi^a 
- n\gamma f^{p-1} {\cal J}^{bc} _{(n-1)} \nabla_b (\gamma f^{-2} \pi_c) \pi^a 
\right. 
\nonumber
\\
&-& \left. n f^{p-2} f' {\cal J}^{ba} _{(n-1)} \pi_b - n \gamma^{-1} f^p f' J^a _{(n-1)} 
+ \gamma f^{-2} \nabla_b (n \gamma^{-1} f^{p+1} {\cal J}^{ba} _{(n-1)}) 
\right. 
\nonumber
\\
&-&\left. \gamma^3 f^{-4} \pi^a \pi_b \nabla_c ( n \gamma^{-1} f^{p+1} {\cal J}^{cb} _{(n-1)} )
\right] \delta \phi - n \gamma^{-1} f^{p+1} {\cal J}^{ab} _{(n-1)} 
\delta (\gamma f^{-2} \pi_b).
\label{eq:Qa-1}
\end{eqnarray}

\noindent
To write down a shorter expression for the eom we first reduce the form
of the divergence term in~(\ref{eq:eom-1}). Repeated application of 
Eqs.~(\ref{eq:nabla-gamma}) and (\ref{eq:id-11}) in (\ref{eq:eom-1}) 
implies that the divergence term reduces to $\nabla_a (-f^{p+1} J^a _{(n)} 
+ n\gamma^{-1} f^p f' J^a _{(n-1)} )$. This expression can be written in 
a somewhat different form when we use the divergence relation for the 
Galileon vectors (\ref{eq:divergence-3}). Hence
\begin{align}
&\nabla_a (-f^{p+1} J^a _{(n)} + n\gamma^{-1} f^p f' J^a _{(n-1)} ) = \gamma L_{n+1}
- (p+1-n\gamma^{-2}) f^{-1} f' L_n  + f^p f' \pi_a J^a _{(n)}
\nonumber
\\
&+ n\gamma f^{p-1} f^{'2}\pi_a J^a _{(n-1)} - 4n \gamma^{-1} f^{p-1} f^{'2} \pi_a J^a _{(n-1)}
+ n\gamma f^{p-2} f' \pi^a \Pi_{ab} J^b _{(n-1)} \nonumber \\ &+ n\gamma^{-2} f^p f' \Pi_{ab}
J^{ab} _{(n-1)}.
\label{eq:id-15}
\end{align}

\noindent
It is straightforward to prove now that by repeatedly 
making the substitution of Eqs. (\ref{eq:Ja}), (\ref{eq:recursive-3}), 
(\ref{eq:conservation-2}), (\ref{eq:id-11}) and (\ref{eq:id-15}) into 
(\ref{eq:eom-1}), the equations of motion reduce to
\begin{eqnarray}
 {\cal E}_n &=& (p-n + \gamma^2) f^{-1}f' L_n - f^p f' \pi_a J^a _{(n)}
 + (1-\gamma^2) f^{-1} f' L_n + 2f^pf' \pi_a J^a _{(n)}
 \nonumber
 \\
 &-& \gamma^{-1} (1-\gamma^2) L_{n+1} - \gamma^{-1} f^{p+1} \Pi_{ab} J^{ab} _{(n)}.
 \nonumber
\end{eqnarray}
Now, by using the divergence identity for the Galileon vectors~(\ref{eq:divergence-3})
we obtain that the equation of motion may be expressed as a conservation law
\begin{equation}
{\cal E}_n = - \nabla_a (f^{p+1} J^a _{(n)}) + \gamma^{-1} f^{p+2} f'
\Tr (J^{ab} _{(n)})=0.
\label{eq:eom-2} 
\end{equation}
In addition, guided by the identity (\ref{eq:id-14}) we may write that
\begin{equation}
{\cal E}_n = \gamma\,L_{n+1}=0.
\label{eq:eom} 
\end{equation}
Clearly, we have only one equation of motion of second order in $\phi$ 
for each value of $n$ and we therefore see that there is only one physical 
degree of freedom. This last compact form is independent of the value of the 
function $f$ and is fully equivalent to the one for the equations of motion 
arising for the Lovelock branes Lagrangians~\cite{bil} when they are expressed 
in the unitary gauge. Certainly, for DBI Galileons the equation of motion 
takes the form
\begin{equation}
{\cal E}_n = - \partial_a  J^a _{(n)} =0.
\label{eq:eom-3} 
\end{equation}

With regards the Noether currents, by similar manipulations on all
the terms in (\ref{eq:Qa-1}) with the use of the identity (\ref{eq:conservation-2}) 
and the recursive relation for the Galileon vectors (\ref{eq:recursive-3}) we obtain 
\begin{equation}
Q^a _{(n)} = \left[ f^{p+1} J^a _{(n)} + n \gamma f^p f' J^a _{(n-1)} \right] \delta \phi
- n f^{p+1} J^{ab} _{(n-1)} \nabla_b \delta \phi.
\label{eq:charges}
\end{equation}
These are the general geometric Noether currents for these Galileon theories associated 
with the Poincar\'e symmetry of the background. Evidently, for the most simple case,
$n=0$, these specialize to $Q^a _{(0)} = f^{p+1} J^a _{(0)} \delta \phi$. Similarly,
note that~(\ref{eq:charges}) specializes to a more simple expression for DBI Galileons
\begin{equation}
 Q^a _{(n)} = J^a _{(n)} \delta \phi - n J^{ab} _{(n-1)} \partial_b \delta \phi.
\label{eq:charges-1}
\end{equation}
It should be noted that  only the $J^{ab} _{(n)}$s and $J^a _{(n)}$s are present 
in (\ref{eq:charges}) which have a nice physical interpretation. These are the 
contractions along the tangent vectors to $\Sigma$~(\ref{eq:tangent}) of the linear 
momentum density, $f^{a\,\mu}$, associated to the Lovelock brane invariants 
(see Eq. (12) in \cite{bil}). The derivation and application of linear momentum 
and angular momentum for this type of branes remains to be developed and discussed 
elsewhere. In fact, in former Galileon field theories an alternative approach for 
obtaining Noether currents and charges was developed in~\cite{nicolis}.

%
\section{Conclusions}
%

In this work, we have first reviewed the flat spacetime Galileon field theories for
any arbitrary dimension in a covariant form. Then we propose a Born-Infeld type framework 
for describing the flat spacetime Galileon Lagrangians. When this is expanded, such 
action becomes a finite series. In fact, the action (\ref{eq:action31}) describes the dynamics of a 
Born-Infeld-Lovelock $p-$brane~\cite{bil} under the unitary gauge where $\phi$ is the 
brane position relative to the foliation i.e., the Goldstone field associated with 
spontaneously broken $N$-dimensional Poincar\'e invariance~\cite{trodden2,trodden}. 
By considering a variational process we obtained the Noether currents associated to 
these models in terms of some tensors, $J^{ab} _{(n)}$, and vectors $J^a _{(n)}$, which 
are in relation with a conserved stress tensor in the Lovelock brane  prescription~\cite{bil}. 
We believe that these tensors will play an important 
role in the Hamiltonian development for the FSGFT which will be reported elsewhere. A 
non-trivial question is how to find the correct couplings $\alpha_n$ leading to a viable 
physical theory but it remains to be explored.
In fact, a related subject has been recently posed by
Hinterbichler et al.~\cite{hinter-6}
Another interesting proposal on this subject 
relating to the unitary analysis of BI gravity theories is developed in Ref.~\cite{tekin2}.
The results here are so far confined to a Minkowski bulk which is a case of sufficient 
complexity but it remains to modify this BI type action in order to consider 
non-trivial interesting maximally symmetric ambient spacetimes in order to include 
curved Galileon field theories defined on the worldvolume. In any case, our approach 
suggests further explorations in order to understand how much this BI-like
structure can be used to analyze issues related to acceleration of branes. 



\section*{Acknowledgments}

We would like to express our thanks to Joel Saavedra for useful discussions.
ER and CC acknowledge partial support from the PRODEP grant UV-CA-320: \'Algebra,
Geometr\'\i a y Gravitaci\'on. Also, ER acknowledges partial support from 
CONACyT grant CB-2009-135297. MC was supported by PUCV through Proyecto DI 
Postdoctorado 2014-2015. This work was partially supported by SNI (M\'exico). MC, CC
and ER acknowledge partial support from CONACyT under grant CB-2012-01-177519-F.
RC also acknowledges support from EDI, COFAA-IPN, SIP-20144150
and SIP-20151031. CC acknowledges partial support by CONACyT Grant 
I0010-2014-02 Estancias Internacionales-233618-C. Also, CC wishes to thank
the Instituto de F\'\i sica-PUCV, Chile, for the warm hospitality where part
of this work was performed.



\appendix


\section{Brane geometry in the unitary gauge}
\label{app:A}

Consider a $N=(p+2)$-dimensional flat Minkowski background spacetime, with isommetry 
algebra the $N$-dimensional Poincar\'e algebra ${\mathfrak{p}}(N-1,1)$. We will assume 
that the bulk adopt a Gaussian normal foliation of the form
\begin{equation}
dS_{p+2} ^2 = f^2(\phi)\,q_{ab}(x^c)\,dx^a dx^b + d\phi^2, 
\label{eq:Gmetric}
\end{equation}
where $q_{ab}$ is a worldvolume metric and $X^{p+1}=\phi$ denotes a Gaussian 
normal transverse coordinate \cite{trodden2}. For the embedding, named 
{\it unitary gauge},
\begin{equation}
 y^\mu = X^\mu (x^a) = \left(
 \begin{array}{c}
x^a 
\\
\phi(x^a)
 \end{array}
\right),
\label{eq:u-gauge}
\end{equation}
the tangent vectors to $\Sigma$ are given by
\begin{equation}
\label{eq:tangent}
e^\mu{}_a = \partial_a X^\mu=
\begin{cases} 
\delta^b{}_a & \mbox{when}\,\, \mu = b,
\\
\nabla_a \phi & \mbox{when}\,\, \mu = p+1.
\end{cases}
\end{equation}
The normal vector $n^\mu$ to $m$ can be obtained from its intrinsic definition, 
namely, $e \cdot n = \eta_{\mu \nu} e^\mu {}_a n^\nu= 0$ and $n\cdot n = \eta_{\mu \nu} 
n^\mu n^\nu = 1$ where $\eta_{\mu \nu}$ is the background spacetime metric
given by (\ref{eq:Gmetric}). Thus,
\begin{equation}
n^\mu = \gamma \left(
\begin{array}{c}
- f^{-2} \nabla^a \phi
\\
1
\end{array}
\right) 
\qquad \mathrm{and} \qquad
n_\mu = \gamma \left(
\begin{array}{c}
- \nabla_a \phi
\\
1
\end{array}
\right),
\label{eq:normal-vectors}
\end{equation}
where $\gamma = \left( 1 + f^{-2}\nabla^a \phi \nabla_a \phi \right)^{-1/2}$. 
From the embedding (\ref{eq:u-gauge}) the induced metric $g_{ab}$ and its 
inverse specialize to
\begin{equation}
g_{ab} = f^2\,q_{ab} + \pi_a \pi_b \qquad \mbox{and} \qquad g^{ab} = f^{-2}\,\left( q^{ab} 
- \gamma^2 \,f^{-2}\,\pi^a \pi^b \right), 
\label{eq:induced-1}
\end{equation}
respectively, where $\nabla_a$ is the covariant derivative compatible with $q_{ab}$
and we have used the notation introduced in Sec. \ref{intro}.
In addition, $\sqrt{-g} = \sqrt{-q}\,\gamma^{-1}\,f^{p+1}$.
Similarly, the extrinsic curvature of the worldvolume is
\begin{equation}
K_{ab} =  \gamma \left( - \Pi_{ab} +2f^{-1} f'\pi_a \pi_b + ff' q_{ab} \right),
\label{eq:KabG}
\end{equation}
where $f' = \partial f / \partial \phi$.
The rate change of $\gamma$ is given by
\begin{equation}
\label{eq:nabla-gamma}
 \nabla_a \gamma = \gamma f^{-1}f' (1-\gamma^2)\pi_a - \gamma^3 f^{-2}
 \pi^b \Pi_{ab}.
\end{equation}
With this expression we obtain a interesting relation involving the matrix~(\ref{eq:Hmatrix})
\begin{equation}
h_{ab} = \gamma f^{-1}f' (q_{ab} - f^{-2} \pi_a \pi_b)
- \nabla_b (\gamma f^{-2}\pi_a).
\label{eq:Hmatrix-2}
\end{equation}

The geometries provided by $G_{\mu \nu}$,
$g_{ab}$ and $q_{ab}$ are closely connected. For instance, the connection symbols
$\mathbf{\Gamma}^\mu _{\alpha\beta}$ associated to $G_{\mu \nu}$ are related to the 
connection coefficients $\gamma^a _{bc}$ associated to $q_{ab}$ as
\begin{eqnarray}
\mathbf{\Gamma}^a _{bc} &=& \gamma^a _{bc},
\\
\mathbf{\Gamma}^{p+1} _{ab} &=& - ff' q_{ab},
\\
\mathbf{\Gamma}^a _{a\,p+1} &=& f^{-1} f' \delta^a{}_b,
\end{eqnarray}
whereas the connection symbols ${\Gamma}^a_{bc}$ associated to $g_{ab}$ are related 
to the connection coefficients $\gamma^a_{bc}$ by means of
\begin{equation}
\label{eq:connections}
{\Gamma}^a _{bc} = \gamma^a _{bc} + \gamma^2 f^{-2} \pi^a\,\Pi_{bc}
+ f^{-1}f' \left( 2\pi_{(b} \delta^a {}_{c)} - \gamma^2 \pi^a q_{bc} 
- 2\gamma^2 \pi^a \pi_b \pi_c \right).
\end{equation}

%
\section{Expansions}
\label{appC}
%

It is useful to consider the first traces of the matrix $h_{ab}$, Eq.~(\ref{eq:Hmatrix})
\begin{eqnarray}
\Tr (h) &=&  \gamma f^{-2} \left\lbrace - [\Pi] + \frac{\gamma^2}{f^{2}}
[\pi^3] + f f' (p+2- \gamma^2) \right\rbrace,
\label{eq:Trh1}
\\
\Tr (h^2) &=& \gamma^2 f^{-4} \left\lbrace [\Pi^2] + 
2\frac{\gamma^2}{f^2} \left( - [\pi^4] + f f' (1-\gamma^2) [\pi^3]\right)
- 2ff' [\Pi] +  \frac{\gamma^4}{f^4} [\pi^3]^2  \right.
\nonumber
\\
&+& \left.  f^2f^{'\,2} (p+4 - 4\gamma^2
+ \gamma^4)  
 \right\rbrace,  
 \label{eq:Trh2}
\\
\Tr (h^3) &=& \gamma^3 f^{-6}\left\lbrace - [\Pi^3] + 3ff' [\Pi^2]
- 3f^2 f^{'\, 2} [\Pi] + 3\gamma^2 f^{-2} [\pi^5] + \gamma^6 f^{-6} [\pi^3]^3
\right.
\nonumber
\\
&-& \left. 3\gamma^2 f^{-1} f' (2- \gamma^2) [\pi^4] 
- 3 \gamma^4 f^{-4}[\pi^4] [\pi^3]  + 3 \gamma^2 f^{'\, 2} (1- 3\gamma^2 + \gamma^4) [\pi^3]
\right.
\nonumber
\\
&+& \left.  3 \gamma^4 f^{-3} f' (1-\gamma^2) [\pi^3]^2 
+ f^3 f^{'\,3} \left[ p + 8 - 12 \gamma^2 + 6 \gamma^4 -
\gamma^6 \right] \right\rbrace.
\label{eq:Trh3}
\end{eqnarray}

\noindent
The characteristic determinant of the matrix ${M}^a{}_b =
\delta^a{}_b + h^a{}_b$ may be expressed as~\cite{lovelock-rund}
\begin{eqnarray}
 \textrm{det} ({M}^a{}_b) &=& 1 + \sum_{s=1} ^{n} \frac{1}{s!}
\delta^{a_1 a_2 a_3 \cdots a_s} _{b_1 b_2 b_3 \cdots b_s} h^{b_1}{}_{a_1}
h^{b_2}{}_{a_2} h^{b_3}{}_{a_3} \cdots h^{b_s}{}_{a_s},
\nonumber
\\
&=& 1 +  \sum_{s=1} ^{n} h_{(s)},
\label{eq:det-2}
\end{eqnarray}
where $ s! h_{(s)} = \delta^{a_1 a_2 a_3 \cdots a_s} _{b_1 b_2 b_3 \cdots b_s} 
h^{b_1}{}_{a_1} h^{b_2}{}_{a_2} h^{b_3}{}_{a_3} \cdots h^{b_s}{}_{a_s}$ denotes 
the determinant of the $s$-rowed minor. These minors can be expressed in terms
of the traces of the $h^a{}_b$ matrix
\begin{eqnarray}
h_{(1)} &=& \Tr (h),
\label{eq:X1}
\\
h_{(2)} &=& \frac{1}{2} \left[ \Tr (h)^2  - \Tr (h^2 ) \right], 
\label{eq:X2}
\\
h_{(3)} &=& \frac{1}{6} \left[ \Tr (h)^3 - 3 \Tr (h^2) \Tr (h) 
+ 2 \Tr (h^3) \right],
\label{eq:X3}
\\
h_{(4)} &=& \frac{1}{24}\left[ \Tr (h)^4 + 8 \Tr (h^3) \Tr (h) 
- 6 \Tr (h^2) \Tr (h)^2 + 3 \Tr (h^2)^2 - 6 \Tr (h^4) \right].
\label{eq:X4}
\end{eqnarray}

\noindent
In some cases it will be useful to obtain
the Taylor expansion of the square root of the characteristic determinant
(\ref{eq:det-2}) which may be obtained by using the well-known expansion
$(1+x)^{1/2} = 1 + \frac{1}{2} x - \frac{1}{8}x^2 + \frac{1}{16} x^3 - 
\cdots$ for $|x|\leq 1$. Hence, 
\begin{equation}
\left[ \textrm{det} \left(\delta^a{}_b + h^a{}_b \right) \right]^{1/2}
= 1 + \frac{1}{2}  \sum_{s=1} ^{n} h_{(s)} - \frac{1}{8} \left(  \sum_{s=1} ^{n} 
h_{(s)}\right)^2  + \frac{1}{16} \left(  \sum_{s=1} ^{n} 
h_{(s)}\right)^3 - \cdots .
\end{equation}
Thus, up to $O(h^6)$ we have
\begin{eqnarray}
\label{eq:expansion}
[\det (\delta^a{}_b  + h^a{}_b)]^{1/2} &=& 1 + \frac{1}{2}\Tr (h) - 
\frac{1}{8} \left[  2 \Tr (h^2) - \Tr (h)^2 \right] 
\nonumber
\\
&+& \frac{1}{48} \left[ 8 \Tr (h^3) - 6 \Tr(h^2)
\Tr(h) +  \Tr(h)^3 \right]
\nonumber
\\
&-& \frac{1}{384} \left[ 48 \Tr (h^4) -32 \Tr(h^3) 
\Tr(h)  
\right. 
\nonumber
\\
&+& \left. 12 \Tr(h^2) \Tr(h)^2 -  12 \Tr(h^2)^2 - \Tr(h)^4 \right] 
\nonumber
\\
&+& \frac{1}{3840} \left[ 384 \Tr (h^5) - 240 \Tr(h^4) 
\Tr(h) 
\right. 
\nonumber
\\
&+& \left. 80 \Tr(h^3) \Tr(h)^2 - 20 \Tr(h^2) \Tr(h)^3 
\right.
\nonumber
\\ 
&+& \left. 60 \Tr(h) 
\Tr(h^2)^2 - 160 \Tr(h^2) \Tr(h^3) + \Tr(h)^5 \right]
\nonumber
\\
&-&  O(h^6).
\end{eqnarray}

%
\section{Galileon tensors and vectors}
\label{app:B}
%

At first few order we have the Galileon tensors
\begin{eqnarray}
J^{ab} _{(0)} &=& f^{-2} \left( q^{ab} - \gamma^2 f^{-2} \pi^a \pi^b \right),
\label{eq:jab-0}
\\
{J}^{ab} _{(1)} &=& \gamma f^{-4} \left\lbrace
\left( -[\Pi] +\frac{\gamma^2}{f^2} [\pi^3]
+ ff' (p+1-\gamma^2) \right)q^{ab} + \Pi^{ab}
\right. 
\nonumber
\\
&+& \left.  \frac{\gamma^2}{f^2} 
\left( [\Pi] - ff'(p+1)\right) \pi^a \pi^b - 2 \frac{\gamma^2}{f^2} \pi^{(a}\Pi^{b)c}\pi_c
\right\rbrace,
\label{eq:jjab-1}
\\
J^{ab} _{(2)} &=& \gamma^2 f^{-6} \left\lbrace
\left( [\Pi]^2 - [\Pi^2] - 2 \frac{\gamma^2}{f^2} \left( [\Pi] [\pi^3] - [\pi^4] \right)
+ 2 \frac{\gamma^2}{f^2} ff' \left( f^2 [\Pi] + p [\pi^3] \right) 
\right. \right.
\nonumber
\\
&-& \left. \left. 2 ff' p[\Pi] + f^2 f^{'2} (p-1) (p+2 - 2 \gamma^2) \right)q^{ab} 
+ 2\Pi^{ac} \Pi^b{}_c 
\right. 
\nonumber
\\
&+& \left.  2 \left( -[\Pi] + \frac{\gamma^2}{f^2} [\pi^3] + ff' (p-\gamma^2) \right)\Pi^{ab}
+  4\frac{\gamma^2}{f^2} \left( [\Pi] - p ff'\right)\pi^{(a} \Pi^{b)c}\pi_c
\right. 
\nonumber
\\
&-& \left. \frac{\gamma^2}{f^2} \left( [\Pi]^2 - [\Pi^2] - 2ff' \, p[\Pi] +
f^2 f^{'2} \, (p-1)(p+2)\right)\pi^a \pi^b
\right. 
\nonumber
\\
&-& \left. 4 \frac{\gamma^2}{f^2} \pi^{(a} \Pi^{b)c} \Pi_{cd} \pi^d 
- 2 \frac{\gamma^2}{f^2} \Pi^a{}_c \Pi^b{}_d \pi^c \pi^d
\right\rbrace.
\label{eq:jjab-2}
\end{eqnarray}

\noindent
Similarly, at first few order we have the Galileon vectors
\begin{eqnarray}
J^a _{(0)} &=& \gamma f^{-2}\,\pi^a,
\label{eq:j0}
\\
J^a _{(1)} &=& \gamma^2 f^{-4} \left\lbrace \left( - [\Pi] + ff' \,p\right)\pi^a 
+ \Pi^a{}_b \pi^b \right\rbrace
\label{eq:j1}
\\
J^a _{(2)} &=& \gamma^3 f^{-6} \left\lbrace \left( [\Pi]^2 - [\Pi^2] - 2 ff' (p-1) [\Pi]
+ f^2 f^{'2} (p-1)p \right)\pi^a 
\right. 
\nonumber
\\
&-& \left. 2 \left( [\Pi] - ff' (p-1) \right) \Pi^a{}_b \pi^b + 2 \Pi^a{}_b \Pi^b{}_c \pi^c 
\right\rbrace.
\label{eq:j2}
\end{eqnarray}

\end{document}